\pdfoutput=1 
\documentclass[%
 reprint,
 amsmath,amssymb,
 aps,
 pra
]{revtex4-2}

\usepackage{natbib}
\usepackage{url}
\usepackage{textcase}

\usepackage{graphicx}
\usepackage{dcolumn}
\usepackage{bm}

\usepackage{color}
\definecolor{blue}{rgb}{0.1,0.1,0.8}
\definecolor{magenta}{rgb}{1,0,1}
\definecolor{red}{rgb}{0.8,0.1,0.1}
\definecolor{green}{rgb}{0.1,0.8,0.1}

\definecolor{blue}{gray}{1.0}

\usepackage[colorlinks=true]{hyperref}

\begin{document}

\renewcommand{\d}{{\rm d}}
\newcommand{\e}{{\rm e}}
\newcommand{\x}{{\rm x}}
\newcommand{\y}{{\rm y}}
\newcommand{\z}{{\rm z}}
\newcommand{\w}{{\rm w}}
\newcommand{\va}{{\sigma}}
\newcommand{\eq}{\begin{equation}}
\newcommand{\eqend}{\end{equation}}
\newcommand{\eqn}[1]{(\ref{#1})}
\newcommand{\Eqn}[1]{Eq.\ (\ref{#1})}
\newcommand{\PD}[2]{\frac{\partial#1}{\partial#2}}
\newcommand{\DD}[2]{\frac{\d#1}{\d#2}}
\newcommand{\pv}{{\bf p}}
\newcommand{\rv}{{\bf r}}
\newcommand{\qv}{{\bf q}}
\newcommand{\vv}{{\bf v}}
\newcommand{\kv}{{\bf k}}
\newcommand{\Pv}{{\bf P}}
\newcommand{\mv}{{\bf m}}
\newcommand{\nv}{{\bf n}}
\newcommand{\Fv}{{\bf F}}
\newcommand{\Rv}{{\bf R}}
\newcommand{\Ev}{{\bf E}}
\newcommand{\Mv}{{\bf M}}
\newcommand{\Dv}{{\bf D}}
\newcommand{\uv}{{\bf u}}
\newcommand{\ev}{{\bf e}}
\newcommand{\sv}{{\bf s}}
\newcommand{\lv}{{\bf l}}
\newcommand{\av}{{\bf a}}
\newcommand{\Lv}{{\bf L}}
\newcommand{\Av}{{\bf A}}
\newcommand{\xv}{{\bf x}}
\newcommand{\Sv}{{\bf S}}
\newcommand{\Bv}{{\bf B}}
\newcommand{\Qv}{{\bf Q}}

\title{Gauge-Invariant Semi-Discrete Wigner Theory}

\author{Mihail Nedjalkov}

\email{mihail.nedialkov@tuwien.ac.at}
\author{Mauro Ballicchia}%
\email{mauro.ballicchia@tuwien.ac.at}
\author{Robert Kosik}%
\email{robert.kosik@tuwien.ac.at}
\affiliation{%
Institute for Microelectronics, Technische Universit{\"a}t Wien, Austria
}%
\author{Josef Weinbub}
\email{josef.weinbub@tuwien.ac.at}
\affiliation{%
 Christian Doppler Laboratory for High Performance TCAD, Institute for Microelectronics, Technische Universit{\"a}t Wien, Austria
}%

\begin{abstract}
A gauge-invariant  Wigner  quantum mechanical theory  is obtained by applying the  Weyl-Stratonovich transform to the von Neumann equation for the density matrix. The transform reduces to the Weyl transform in the electrostatic limit, when the vector potential and thus the magnetic field are zero. Both cases involve a center-of-mass transform  followed by a  Fourier integral on the relative coordinate introducing the momentum variable. The latter is continuous if the limits of the integral are infinite or, equivalently, the  coherence length is infinite. However, the quantum theory involves Fourier transforms of the electromagnetic field components, which imposes conditions on their behavior at infinity. Conversely, quantum systems are bounded and often very small, as is, for instance, the case in modern nanoelectronics. This  implies  a finite coherence length, which avoids the need to regularize non-converging Fourier integrals. Accordingly, the momentum space becomes discrete, giving rise to momentum quantization and to a semi-discrete gauge-invariant Wigner equation. To gain insights into the peculiarities of this theory one needs to analyze the equation for specific electromagnetic conditions. We  derive the evolution equation for the linear electromagnetic case and  show that  it significantly simplifies for a limit dictated by the long coherence length behavior, which involves momentum derivatives. In the discrete momentum picture these derivatives are presented by finite difference quantities which, together with further approximations, allow to develop a  computationally feasible model that offers physical insights into the involved quantum processes. In particular, a Fredholm integral equation  of the second kind is obtained, where the "power" of the kernel components, measuring their rate of modification of the quantum evolution, can be evaluated.
\end{abstract}

\maketitle

\section{\label{sec:Int}Introduction}
The description of the quantum evolution of charged particles in an electromagnetic (EM) medium is a fundamental problem in many areas, particularly in nanoelectronics~\cite{Novakovic2011,DuqueGomez2012,Nedjalkov2015,Iafrate2017,Bellentani2019,PRBMagnetic,Ferry2020,MixiBook2021,Iafrate2021,Fatima2021,Cepellotti2021,SST_Ferry_2022,Iafrate2022}. 
Several approaches with different properties are available and actively developed to describe the quantum processes  (for a recent review of relevant computational methods see~\cite{Weinbub_2022}). In particular, the Schr{\"o}dinger equation and the non-equilibrium Green's functions formalism rely on the boundary conditions to enable analysis in terms of eigenstates and eigenvalues, while the density matrix and the Wigner function approach need the initial condition of the considered system to describe the future evolution~\cite{SST_Ferry_2022}. 

The Wigner formalism provides a very intuitive formulation of quantum mechanics,  maintaining many classical concepts
and notions such as the phase space and physical observables  represented by the  same functions of position and  momentum as the classical counterparts~\cite{Wigner1932, Ferry_Mixi_book_2018,Weinbub2018}.
The Wigner function is a real quantity and is used to calculate physical averages in the same manner as with the classical distribution function.  
Furthermore, coherence breaking processes can be included in a straightforward manner by using scattering functions in the governing Wigner equation in the same way as in the classical counterpart - the Boltzmann equation. 
Here, we focus on the purely coherent Wigner equation~\footnote{Historically the Wigner equation has been derived from the Schr{\"o}dinger equation or for mixed states from the von Neumann equation for the density matrix. It is interesting to note that  the proof that the Wigner formalism is an autonomous theory  came two decades later: The works of Moyal  and Groenewold established  the formalism  as a self-contained formulation of quantum mechanics~\cite{Moyal49, Groenewold46}. In this way, the Wigner formalism is fully equivalent to operator mechanics which can be derived on top of the formalism~\cite{Dias2004110}. The involved Moyal bracket and star-product, however, offer a high level of abstraction, so that the intuitive historical approach,  which works in terms of center-of-mass and Fourier transforms (FTs), called Weyl transform, remains widespread.}.

In the electrostatic limit and considering a zero vector potential gauge the transport problem can be conveniently formulated with the help of the electric potential $\phi$. The Weyl transform of the von Neumann equation for the density matrix
$\rho(r_1,r_2,t)$ defines two central  quantities, the Wigner function 
\begin{equation}
f_w(p,x,t)=C\int ds
\rho(x+\frac{s}2,x-\frac{s}2,t)e^{-\frac i\hbar sp}
\label{WT}
\end{equation}
and the Wigner potential
\begin{equation}
V_w(p,x)=\frac{C}{i\hbar}\int ds
\left(
V(x-\frac{s}2)-V(x+\frac{s}2)
\right)
e^{-\frac i\hbar sp} \,.
\label{WP}
\end{equation}
Here, the two positions $r_1$ and $r_2$ are expressed through the center-of-mass coordinates $x,s$ and $V=e\phi$ is the energy of the electron due to the electric potential.
$C$
is a normalization constant which depends on the limits of integration and the dimensionality of the task.
For a one-dimensional (1D)  $s$ and integration spanned over the whole space (infinite coherence length $L$)  $C=\frac{1}{(2\pi\hbar)}$.
Then $x$ and $p$ form a continuous phase space.
In this picture the canonical and the kinetic momenta coincide, i.e.,  the integrals $\int f_wdx=|\psi(p)|^2$,  $\int f_wdp=|\psi(x)|^2$ (where $\psi$ is the wave function) give the distributions of the eigenvalues of the conjugate momentum and position operators $\hat p=-i\hbar\nabla$ and $\hat x$, while the phase space integral of $pf_w$ divided by the mass $m$ gives the mean  velocity.
The physical quantities are represented by the same dynamical functions
inherent to classical mechanics, which are often devised in terms of position and velocity.

The presence of a magnetic field enormously complicates the standard Wigner equation (see (5.22) in~\cite{Nedjalkov2011}): The description becomes multi-dimensional, and the canonical, $\pv$, and the kinetic, $\Pv$, momenta differ by the vector potential $\Av$ via $\pv=\Pv+e\Av(\xv)$. One can still use the Weyl transform \eqn{WT} to derive a quantum theory in phase space.
However, this introduces a dependence on the  gauge via $\pv$ which depends on the vector potential.
Then the equation for the Wigner distribution function describing the electron state changes with any choice of a new gauge. 
The  same holds for dynamical functions, which are defined in terms of the kinetic momentum.
The other option is to modify the Weyl transform: 
In principle, alternative formulations for quantum mechanics in phase space have already been introduced. 

For instance, in a comprehensive relativistic (quantum electrodynamics) study
the evolution equations were derived for fermion and photon many-body
Wigner operators and analyzed with respect to the interplay between the space-time Lorentz and the electromagnetic gauge transforms within the context of a Lorentz-covariant and
gauge-invariant quantum transport theory~\cite{Vasak1987}. 
The non-relativistic limit which is appropriate for condensed matter physics
was developed in~\cite{Serimaa86} (also cited in~\cite{Vasak1987}).
In this work, a single particle gauge-invariant Wigner operator  and a gauge-independent Wigner function together with the corresponding operator
equation of motion was developed. The latter depends only on
the EM forces and is thus independent of the choice of gauge.
The meaning of "gauge invariance" used in this work follows the broadly accepted view~\cite{Serimaa86,Haug2008}.
A peculiarity of the derived  evolution equation (see also \cite{Levanda01})
is that the spatial arguments of the EM fields are replaced by operators.
This enables an elegant formulation of the gauge-invariant evolution
equation, but makes the numerical treatment enormously difficult. 
In our previous work \cite{PRBMagnetic}, we present an effort to reformulate the equation
in terms of continuous phase space mathematical operations which are independent of the shape of the EM fields.
This, however, imposes some special conditions on the behavior of the EM fields
at infinity, which motivates the semi-discrete formulations presented in this work.
It is interesting to note that \cite{Serimaa86,Vasak1987,Levanda01,Haug2008,PRBMagnetic} provide  self-contained derivations of the transport theory. Furthermore, they are unified around the heuristic idea of replacing the
adjoint momentum with the kinetic momentum: They are "uniquely determined by requiring that the
momentum variable corresponds to the kinetic momentum"~\cite{Vasak1987}. 
This idea can be traced back to the work of Stratonovich~\cite{Stratonovich56}, who derived the generalization of the Weyl transform giving rise to the kinetic momentum.
Besides, the dynamical functions preserve their classical form independent of the chosen gauge. The derivation of the Weyl-Stratonovich transform (WST)~\cite{PRBMagnetic},
\begin{eqnarray}
\label{WWS}
&&f_w(\Pv,\xv)=\\
&&\int
\frac{d\sv}{(2\pi\hbar)^3}
e^{-\frac{ i}{\hbar} \sv\cdot\Pv}
e^{ -\frac{i}{\hbar }
\frac e2 \sv\cdot\int\limits_{-1}^{1}
 d\tau
\Av(  \xv+\frac{\sv\tau}{ 2})
}
\rho(\xv+\frac{\sv}{2},\xv-\frac{\sv}{2} ) ,
\nonumber
\end{eqnarray}
is based on an analysis of the way the mean values of  products of the components of the
canonical momentum change after a canonical transform~\cite{Stratonovich56}. 
It is interesting to note that the evolution equation for \eqn{WWS} was not presented in the original manuscript of Stratonovich. However, alternative versions of this equation have been presented in several  subsequent works~\cite{KuboWmag64, Serimaa86, Serimaa87, Levanda01, Materdey03n1, Materdey03n2}. In general, these works use pseudo-differential operators comprised of
physical functions, such as EM fields,  where the  arguments contain not only the regular  phase space variables  but also differential operations with these variables. 
This means that equations of this type 
are implicit with respect to the mathematical appearance, in
particular, the order of the differential part depends on the way these functions change with the physical environment. As hinted above, in previous work, we derived a version of the  Wigner equation from the
evolution equation for the  density matrix
formulated with the help of scalar and vector potentials, corresponding to general  EM conditions~\cite{PRBMagnetic}. The  WST is then applied
to introduce  the kinetic momentum and thus a gauge invariance:
Indeed the obtained equation depends only on the EM forces.

For homogeneous magnetic fields inhomogeneous terms vanish
and the equation reduces to its electrostatic limit form with an additional term
accounting for the acceleration due to the magnetic field~\footnote{
This equation is already numerically feasible, e.g., the effect of the magnetic field on typical quantum processes has been analyzed~\cite{ieeeNano2020}. Furthermore, if a constant electric field is assumed, the equation reduces to the well-known ballistic Boltzmann equation with the Lorentz force driving the electron distribution over  Newtonian trajectories.}.
When considering inhomogeneous magnetic terms, the first problem is that they pose serious mathematical
challenges because of multi-dimensional integrals.
A further problem comes from the infinite limits of the Fourier integral in (\ref{WT}): 
The Wigner function is well-defined in these limits, because $\rho\in{\cal L}_2$~\footnote{This  can be seen by a direct application of the Cauchy–Schwarz inequality first to a pure state, $\rho=\psi\psi^*$, and then generalized for mixed states.}.
However, this does not hold
for \eqn{WP}, e.g., in the case of an electrostatic potential step.
The problem can be resolved by introducing the \emph{heresy} of generalized functions~\cite{Vladimirov84} into the Wigner picture~\cite{Nedjlkov97_P9}. 
However, generalized functions involve special limits~\cite{Vladimirov84} which require analytical approaches and can turn into operators  as discussed in  Appendix~\ref{AAC}.
 The usual numerical procedure of  discretization by presenting integrals by Riemann sums fails. 
On the contrary, the parent equation for the
density matrix, or the $\sigma$-equation, is suitable for such a treatment~\cite{Kosik_2021}~\footnote{The sigma function is the inverse FT of the Wigner function in the non-spatial coordinate. The resulting equation is mathematically close to the Liouville-von Neumann equation but inherits a physical interpretation from the Wigner function formalism~\cite{Schulz_2019,Kosik_2021}.}.
This shows that the problem is introduced by the properties of the mathematical transforms at infinity and not by the involved physics.
Indeed, a physically admissible Schr{\"o}dinger state   lies in the domain of the self-adjoint momentum operator and thus vanishes at infinity, $lim_{|x|\to\infty}\psi(x)=0$.
This property implies an alternative "bounded domain" approach to the formalism: 
For convenience we consider states $\psi$ which evolve in a  bounded domain $\Omega$.
The condition $\psi=0$ outside  $\Omega$ characterizes  the physical settings in a broad class of problems such as systems initially restricted by  potentials. If such a  system  opens at  a given time,  the domain can extend, but remains bounded for  finite evolution intervals~\footnote{Indeed the appearance of a finite density at infinity indicates state components with infinite speed/kinetic energies.}.
A continuous state which evolves   in $\Omega$ has a well-defined discrete Fourier image $f_n$ in a discrete momentum space~\footnote{The coherence length $\Lv$ of the FT must be greater than $\Omega$ because the function (recovered by $f_n$) becomes periodic with $\Lv$: Strictly speaking, if we wish to guarantee that the function is zero everywhere outside $\Omega$, then $\Lv$ should become infinity. However, we are interested in the  domain $\Omega$, where the function coincides with the state.}.
This suggests to develop a discrete momentum  EM Wigner theory which will be particularly suitable for numerical treatment of the inhomogeneous magnetic terms.

In this work, we formulate the general form of the discrete momentum EM Wigner formalism (Section~\ref{sec:level1}). It is formulated with the help of quantities defined by the FT of the EM field components, which remain well-defined for a finite coherence length.  
Within the established formalism we focus on the special case of linear EM fields (corresponding to the first-order terms of their Taylor expansion) and derive the corresponding Wigner evolution equation (Section~\ref{Appl}). 
Finally, we use it to develop an approximate integral form suitable for numerical solution approaches.

\section{\label{sec:level1}The Semi-discrete Wigner Formalism}
\subsection{\label{sec:level2}Discrete Momentum Wigner Function}
We consider a system described by a density matrix
$\rho(\rv_1,\rv_2)$, which becomes zero outside a given  domain $\Omega$ with dimensions $(0,\Lv/2)$, where the components of $\Lv/2$ along the coordinate axes define the extent of our system,  $0<\rv_1,\rv_2<\Lv/2$.
For the center-of-mass variables $\xv=\frac{\rv_1+\rv_2}{2},\quad \sv=\rv_1-\rv_2$, this condition becomes 
\eq
0<\xv<\frac\Lv2 \,,
\qquad
-\frac\Lv2<\sv<\frac\Lv2 \,.
\label{bounds}
\eqend
The FT of a continuous function $f(\sv)$ is defined as 
\eq
f_\nv=\frac{1}{\Lv} \int_{-\Lv/2}^{\Lv/2}d\sv e^{-i\nv\Delta\kv\sv}f(\sv)  \,,
\qquad
f(\sv)=\sum_{\nv=-\infty}^{\infty}e^{i\nv\Delta\kv\sv}f_\nv  \, ,
\label{FT}
\eqend
where $\nv\Delta\kv$ denotes the vector with components $n_i\Delta k_i$ along the coordinate axes $i=1,2,3$. The uniqueness of the decomposition follows from the condition for orthonormality.
The completeness relation follows from \eqn{FT}
\eq
 \Delta\kv=2\pi/\Lv \, ,
 \qquad
\frac{1}{\Lv}
\sum_{\mv=-\infty}^{\infty}
e^{i\mv\Delta\kv(\sv-\sv')}
=
\delta(\sv-\sv') \, ,
\label{complete}
\eqend
 where $\Lv$ determines the momentum discretization.
We continue by using the the momentum variable $\Pv=\hbar\kv$.
The definition of our Wigner function is then
\begin{eqnarray}
\label{WWSdis}
f_w(\Pv_{\mv},\xv)=&&
\int_{-\Lv/2}^{\Lv/2}
\frac{d\sv}{\Lv}
e^{-\frac{ i}{\hbar} \sv\cdot\Pv_{\mv}}\\
&&e^{ -\frac{i}{\hbar }
\frac e2 \sv\cdot\int\limits_{-1}^{1}
 d\tau
\Av(  \xv+\frac{\sv\tau}{ 2})
}
\rho(\xv+\frac{\sv}{2},\xv-\frac{\sv}{2} ) \, ,
\nonumber
\end{eqnarray}
where  $\Pv_{\mv}=\mv\Pv$ is a vector with components 
${(m_x\Delta p_x, m_y\Delta p_y,m_z\Delta p_z)}$. 
In this way  $f_w( \Pv_{\mv},\xv ,t)$ is continuous
with respect to $\xv$ and discrete with respect to  three integer numbers $\mv$. 
The time variable remains implicit. For a better transparency we will interchangeably use
both notations $\Pv_{\mv}$ and $\mv\Pv$, which puts the focus on the physical or mathematical
aspects, respectively.

\subsection{Discrete Momentum Evolution Equation}
We begin to reformulate the von-Neumann
equation for the density matrix in center-of-mass
coordinates~\cite{PRBMagnetic}:
\begin{widetext}
\begin{eqnarray}
&&\frac1{2mi\hbar}
\left[\sum_l
2\left(i\hbar \frac{\partial}{\partial x_l} 
+ eA_l(\xv+\frac\sv 2)- eA_l(\xv-\frac\sv 2)
\right)
\left(i\hbar\frac{\partial}{\partial s_l} + \frac e2 A_l(\xv-\frac\sv 2)+ \frac e2 A_l(\xv+\frac\sv 2)\right)
\right]=
\nonumber \\
&&
-\frac{1}{i\hbar}\left( V(\xv+\frac\sv 2)-V(\xv-\frac\sv 2)\right)\rho(\xv+\frac\sv 2,\xv-\frac\sv 2)
+\frac{\partial\rho(\xv+\frac\sv 2,\xv-\frac\sv 2)}{\partial t}
\label{generic2}
\end{eqnarray}
We multiply  equation   (\ref{generic2}) by the exponent factor
in (\ref{WWSdis}) and integrate over $\sv$. Then the exponent factor must be shifted to the right, next to the density matrix. We first consider the product of the two brackets:
In the electrostatic limit, $\Av=0$, this shift is straightforward, but for general EM fields the differential operators in the brackets do not commute with $\Av$. 
Fortunately, all related transforms from the continuous derivation in~\cite{PRBMagnetic} apply also in the discrete case:
\begin{eqnarray}
  && {\cal D}=
\int\limits_{-\Lv/2}^{\Lv/2}\frac{d\sv}{\Lv}
\left\{-
\frac{\Pv_{\Mv}}{m}\cdot
\frac{\partial}{\partial \xv} 
-
\frac 1m
\frac e2 \int\limits_{-1}^{1} d\tau
\frac\tau 2
\left[\sv\times\Bv(  \xv+\frac{\sv\tau}{ 2})\right]
\cdot\frac{\partial}{\partial \xv} 
+\frac {e}{2i\hbar}
\int\limits_{-1}^{1} 
d\tau
\left[\sv\times\Bv(  \xv+\frac{\sv\tau}{ 2})\right]\cdot\frac{\Pv_{\Mv}}{m}
\right.
\label{De}
\\
&&
\left.
+
\frac {e^2}{4mi\hbar}
\int\limits_{-1}^{1} 
\int\limits_{-1}^{1} d\tau
d\eta\frac\tau 2
\left[\left(\sv\times\Bv(  \xv+\frac{\sv\eta}{ 2})\right)\cdot
\left(\sv\times\Bv(  \xv+\frac{\sv\tau}{ 2})\right)\right]
\right\}
e^{ -\frac{i}{\hbar }\sv\cdot\left(\Pv_{\Mv}+
\frac e2 \int\limits_{-1}^{1}  d\tau \Av(  \xv+\frac{\sv\tau}{ 2})\right)}
\rho(\xv+\frac\sv 2,\xv-\frac\sv 2)
\nonumber
\end{eqnarray}
\end{widetext}
The term after the curly brackets gives the Wigner function \eqn{WWSdis} when integrated on $\sv$.
Thus we need to decouple it from  the expression in the curly brackets, which is done with the help of the completeness relation \eqn{complete}: 
The term with the exponent is introduced into \eqn{De} and as it results in a delta function we can change $\sv$ to $\sv'$ in the curly brackets and then integrate over   $\sv'$ to recover the value of ${\cal D}$. 
This leads to  a separate FT of the consecutive terms in the curly brackets. 
The square brackets show that we actually need the function
\eq
H^F(\xv,\mv,\tau)=
\int\limits_{-\Lv/2}^{\Lv/2}
\frac{d\sv'}{\Lv}
e^{-\frac i\hbar\mv\Delta\pv \sv'}\left[\sv'\times\Bv(  \xv+\frac{\sv'\tau}{ 2})\right] \,. 
\label{HF}
\eqend
Indeed the integral
\begin{eqnarray}
&&I^F(\xv,\mv,\tau)=
\label{IF}\\
&&\int\limits_{-\Lv/2}^{\Lv/2}
\frac{d\sv'}{\Lv}
e^{-\frac i\hbar\mv\Delta\pv \sv'}
\left(\sv'\times\Bv(  \xv+\frac{\sv'\eta}{ 2})\right)\cdot
\left(\sv'\times\Bv(  \xv+\frac{\sv'\tau}{ 2})\right)
\nonumber
\end{eqnarray}
can be expressed via the convolution
$I^F(\xv,\mv,\eta,\tau)=H^F(\xv,\mv,\eta)*H^F(\xv,\mv,\tau)$.

Finally we obtain for ${\cal D}$:
\begin{widetext}
\begin{eqnarray}
&&
\hskip-1cm
{\cal D}=
\sum_{\mv=-\infty}^{\infty}
\left\{-\delta_{\mv,0}
\frac{\Pv_{\Mv}}{m}\cdot
\frac{\partial}{\partial \xv} 
-
\frac e{2m}
\int\limits_{-1}^{1} d\tau
\frac\tau 2
H^{F}(\xv,\mv,\tau)\cdot
\frac{\partial}{\partial \xv} 
+
\frac {e}{2i\hbar}
\int\limits_{-1}^{1} 
d\tau
H^{F}(\xv,\mv,\tau)\cdot\frac{\Pv_{\Mv}}{m}
\right.
\nonumber\\
&&
\left.+
\frac {e^2}{4mi\hbar}
\int\limits_{-1}^{1} 
\int\limits_{-1}^{1} d\tau
d\eta \frac\tau 2
I^F(\xv,\mv,\tau,\eta)
\right\}
f_w\bigl(\Pv_{\Mv-\mv},\xv\bigr) 
\label{Ddetailed}
\end{eqnarray}
\end{widetext}

The right-hand side ${\cal T}$
of \eqn{generic2} is handled in the same way; the first term can be directly processed: 
\begin{eqnarray}
&&
\int\limits_{-\Lv/2}^{\Lv/2}
e^{ -\frac{i}{\hbar }\sv\cdot\left(\Pv_\Mv
+\frac e2 \int\limits_{-1}^{1} d\tau \Av(  \xv+\frac{\sv\tau}{ 2})\right)}
\left( V(\xv-\frac\sv 2)-V(\xv+\frac\sv 2)\right)
\nonumber\\&&
\rho(\xv+\frac\sv 2,\xv-\frac\sv 2) \frac{d\sv}{i\hbar\Lv}
=\!
\sum_{\mv=-\infty}^{\infty}\!
V_w(\mv,\xv)f_w\bigl(\Pv_{\Mv-\mv},\xv\bigr) 
\label{calP}
\end{eqnarray}
We recognize the Wigner potential \eqn{WP} in the first line of \eqn{calP}, now formulated in the discrete  momentum space. The second term on the right-hand side of \eqn{generic2} can be evaluated by using the time derivative of  (\ref{WWSdis}):

\begin{eqnarray}
&&
\int\limits_{-\Lv/2}^{\Lv/2}
\frac{d\sv}{\Lv}
e^{ -\frac{i}{\hbar }\sv\cdot\left(\Pv_\Mv+
\frac e2 \int\limits_{-1}^{1} d\tau \Av(  \xv+\frac{\sv\tau}{ 2})\right)}
\frac{\partial\rho(\xv+\frac\sv 2,\xv-\frac\sv 2)}{\partial t}
=
\nonumber\\
&& \frac{\partial}{\partial t}f_w(\Pv_\Mv,\xv)
-
\frac {e}{i2\hbar\Lv }
\int\limits_{-1}^{1} d\tau
\int\limits_{-\Lv/2}^{\Lv/2}d\sv\, 
\sv\cdot
  \frac{\partial  \Av(  \xv+ \frac{\sv\tau}{2})}{\partial t}
\nonumber\\
&&
e^{ -\frac{i}{\hbar }\sv\cdot\left(\Pv_\Mv\Delta\pv+
\frac e2 \int\limits_{-1}^{1} d\tau \Av(  \xv+\frac{\sv\tau}{ 2})\right)}
\rho(\xv+\frac\sv 2,\xv-\frac\sv 2)
\label{rhs}
\end{eqnarray}
 The identity
$ -\frac{\partial  \Av}{\partial t}=\nabla  \phi+\Ev$
can be used to eliminate $\partial\Av/\partial t$ from \eqn{rhs}.
The contribution from the scalar potential can be directly integrated over $\tau$:
\begin{eqnarray}
\frac {e}{ 2}
\int\limits_{-1}^{1} d\tau
\sv\cdot
\frac{\partial}{\partial\xv}  \phi(  \xv+ \frac{\sv\tau}{2})
=\label{phiV}
V(  \xv+ \frac{\sv}{2})- V(  \xv- \frac{\sv}{2})
\nonumber
\end{eqnarray}
Consequently, the contribution from $\phi$ cancels the Wigner potential term \eqn{calP}.

Therefore, only the electric field $\Ev$  contributes, so that 
\begin{eqnarray}
&&
{\cal T}=\frac{\partial}{\partial t}f_w(\Pv_\Mv,\xv) -
\label{tau}\\&&
\frac {e}{2i\hbar}
\sum_{\mv=-\infty}^{\infty}
\int\limits_{-1}^{1} d\tau
D^F(\xv,\mv,\tau)
f_w\bigl(\Pv_{\Mv-\mv},\xv\bigr)
\nonumber
\end{eqnarray}
with
\eq
D^F(\xv,\mv,\tau)
= -
\int\limits_{-\Lv/2}^{\Lv/2}
\frac{d\sv'}{\Lv}
e^{-\frac i\hbar\mv\Delta\pv \sv'}\left(
\sv'\cdot\Ev(  \xv+ \frac{\sv'\tau}{2}) 
\right) \,.
\label{DF}
\eqend
The evolution equation for $f_w$, given by the equality 
\eq
{\cal D}={\cal T} \, ,
\label{main}
\eqend
is finally obtained:
\begin{widetext}
\begin{eqnarray}
&&
\left(
\frac{\partial}{\partial t}
+
\frac{\Pv_{\Mv}}{m}\cdot
\frac{\partial}{\partial \xv} 
\right)
f_w(\Pv_\Mv,\xv) =
\sum_{\mv=-\infty}^{\infty}
\left\{
\frac {e}{2i\hbar}
\int\limits_{-1}^{1} d\tau
D^F(\xv,\mv,\tau)
-
\frac e{2m}
\int\limits_{-1}^{1} d\tau
\frac\tau 2
H^{F}(\xv,\mv,\tau)\cdot
\frac{\partial}{\partial \xv} 
\right.
\nonumber
\\
&&\left.
+\frac {e}{2i\hbar}
\int\limits_{-1}^{1} 
d\tau
H^{F}(\xv,\mv,\tau)\cdot\frac{\Pv_{\Mv}}{m}
+
\frac {e^2}{4mi\hbar}
\int\limits_{-1}^{1} 
\int\limits_{-1}^{1} d\tau
d\eta \frac\tau 2
I^F(\xv,\mv,\tau,\eta)
\right\}
f_w\bigl(\Pv_{\Mv-\mv},\xv\bigr)
\label{IWE}
\end{eqnarray}
\end{widetext}

\subsection{\label{PoE}Properties of the Equation}
The obtained integro-differential equation \eqn{IWE} has three kernels, $D^F$, $H^F$, and $I^F$, which depend on $\Ev$ and $\Bv$. Therein lies the gauge-invariance of the equation as these quantities are independent of the choice of the gauge. 

Equations \eqn{HF}, \eqn{IF}, and \eqn{DF} are obtained by well-defined mathematical operations due to the finite domain of the integration.
However, the derivation of \eqn{IWE} doesn't guarantee that the $\Lv\to\infty$ / continuous momentum limit of the equation exists.
Indeed, already a constant electric field in
\eqn{DF} gives rise to terms 
\eq
D(\mv)\propto (-1)^\mv/(\mv\Delta \pv)
\label{harmonic}
\eqend
which diverge at this limit.
The continuous FT raises  the need to restrict the behavior of the EM fields when $\sv\to\infty$ in order to ensure convergence, e.g.,  $\sv\Ev$ and $\sv\Bv$  must be absolutely integrable, or alternatively employ the formalism  of generalized functions.
The latter is based on the following: The alternating harmonic series \eqn{harmonic}
already challenges  the discrete momentum case, because the reordering of the terms of the  series can make it to converge to any number. Fortunately, these terms are multiplied by the "good" function $f_w$ and  this is what  regularizes the sum in \eqn{IWE}. 
Similarly to the discrete case, the "good" behavior of $f_w$ at infinity allows to introduce a factor $e^{-\alpha|\sv|}$  in the corresponding integrals and then to consider the limit $\alpha\to 0$~\cite{Nedjlkov97_P9}.
However, as discussed, the application of this formalism is not convenient from a practical point of view~\footnote{The use of a  bounded domain of integration  has also a physical motivation. It reflects  the fact that the density matrix becomes zero outside some bounds $\Omega\le \Lv$.  Namely, at  a given evolution time $f(\sv)=0$ outside a given $\Omega$, a choice of a larger  $\Lv$  would change the involved frequencies, but not the limits of integration for the coefficients $f_\mv$ in \eqn{FT}.}.

Another feature of \eqn{IWE} is that the second and forth term in the curly brackets vanish for homogeneous magnetic fields. This greatly simplifies the equation, which in the continuous version can be reformulated in terms of the Lorentz force. The equation is convenient for numerical implementation: A  model has been developed and applied to study magnetoresistance~\cite{PRBMagnetic}.
However, the vanishing terms have not been investigated with respect to inhomogeneous magnetic fields (i.e., spatial variations of $\Bv$). 
In this case and considering "small" dimensions (e.g., nanometer scales), the linear term in the Taylor expansion becomes physically relevant.

Equation \eqn{IWE} can be further reformulated: The idea is to use the Fourier images of $\Ev$ and $\Bv$ in \eqn{HF}, \eqn{IF}, and \eqn{DF} and to solve the explicit integrals involving the exponents. This will be done in the next section, where we explore the equation for the case of a  linear magnetic field. 
The electric field is also assumed linear, which corresponds to the  case of an applied bias.

\section{\label{Appl} Linear Electromagnetic Fields}
In this case most of the mathematical operation can be carried out analytically, which greatly simplifies the equation.  The latter is further approximated  to reduce the complexity towards a numerically feasible model. We first neglect one of the magnetic terms in ${\cal D}$ and then analyze the   continuous limit of the equation. 
These steps including further simplifications provide a convenient starting point to derive the corresponding integral form.

For simplicity a 2D evolution is considered and an arrow is used to denote vectors, for example,
$\frac{\partial}{\partial \xv} =
\overrightarrow{\left(\frac{\partial}{\partial x}, \frac{\partial}{\partial y} \right)} $.
The problem corresponds to  an electron state evolving in the $x,y$ plane   under the action of a a magnetic field  $\Bv=\overrightarrow{(0,0,B(y))}$ normal to the plane. The field is inhomogeneous in the $y$ direction,  $B(y)=B_0+B_1y$.
The electric field is $\Ev(x,y)=(E_xx,E_yy)$ and is determined by an applied bias on the boundaries.

\subsection{Finite Coherence Length}
Here, we derive the evolution equation corresponding to  \eqn{main}.  
We first consider the particular expression for ${\cal T}$:
\begin{eqnarray}
&&
{\cal T}=\frac{\partial}{\partial t}f_w(\Pv_\Mv,\xv) -\frac {e}{2i\hbar}
\int\limits_{-\Lv/2}^{\Lv/2}
\frac{d\sv}{\Lv}
\sum\limits_{\mv=-\infty}^{\infty}
\label{taulin}\\&&
\int\limits_{-1}^{1} d\tau
e^{-\frac i\hbar\mv\Delta\pv \sv}\left(
\sv\cdot\Ev(  \xv+ \frac{\sv\tau}{2})
\right)
f_w\bigl(\Pv_{\Mv-\mv},\xv\bigr)
\nonumber
\end{eqnarray}
The $\tau$ and $\sv$ integrals can be carried out analytically.
The terms which are linear in $\tau$ give zero contribution due to the symmetric bounds, so that the
second term in \eqn{taulin} is proportional to

\eq
\int\limits_{-\Lv/2}^{\Lv/2}
\frac{d\sv}{\Lv}
\sum\limits_{\mv=-\infty}^{\infty}
e^{-\frac i\hbar\mv\Delta\pv \sv}
\left(
E_xxs_x+
E_yys_y
\right)f_w(\Pv_{\Mv-\mv}, \xv) \, .
\label{taucal}
\eqend
The evaluation  of the integrals on $s_x$ and $s_y$ can be carried out  using integration by parts.
The obtained expression for ${\cal T}$ is given in Appendix~\ref{AA0}.

Similarly, the $\tau$ integral in the expression for ${\cal D}$ can be directly evaluated. This gives rise to terms containing higher order powers of the components of $\sv$:
\begin{widetext}
\begin{eqnarray}
  && {\cal D}=
\int\limits_{-\Lv/2}^{\Lv/2}\frac{d\sv}{\Lv}
 \sum_{\mv=-\infty}^{\infty}
e^{-\frac i\hbar \mv \Delta\pv \sv}
\left[-
\frac{\Pv_\Mv}{m}\cdot
\frac{\partial}{\partial \xv} 
+
\frac{eB(y)}{i\hbar}
\frac{\Pv_\Mv}{m}\cdot
\overrightarrow{
\left(s_y, -s_x\right)
}
\right.
\nonumber\\&&
-
\frac {B_1}m
\frac e{12}
\overrightarrow{
\left(s_y^2, 
-s_ys_x\right)}
\cdot\frac{\partial}{\partial \xv} 
\left.
+
\frac {e^2B_1B(y)}{12mi\hbar}
\left(s_y^3+ 
s_x^2 s_y\right)
\right]
f_w\bigl(\Pv_{\Mv-\mv},\xv\bigr) 
\label{Deafter}
\end{eqnarray}
\end{widetext}
This makes the evaluation of the $\sv$ integral a cumbersome but straightforward process,
based on applying the integration by parts rule several times.
At this stage we neglect the quadratic magnetic field term in ${\cal D}$:
For a broad class of evolution problems on the nanometer scale the last term in
the square brackets in \eqn{Deafter} is one or more orders of magnitude less than the
rest of the terms (an example is given in Appendix~\ref{AAA}) and is thus neglected in the following.
The resulting expression for ${\cal D}$, combined with the result \eqn{tauend},
gives rise to the component-like form of equation \eqn{main} given in Appendix~\ref{AAB}.
In what follows, we present a vector form of the equation, showing how the action of the 
Lorentz force, enclosed in the square brackets,  is generalized
in the quantum case.
\begin{widetext}
\begin{eqnarray} 
&&\left(\frac{\partial}{\partial t}
+
\frac{\Pv_\Mv}{m}\cdot
\frac{\partial}{\partial \xv}\right)f_w(\Pv_\Mv,\mv)
=
{\sum\limits_{\mv=-\infty}^{\infty}}
\left\{\phantom\int\right.
-e
\left[ \phantom{\frac11}
\Ev(\xv)+\frac{\Pv_\Mv}{m}\times\Bv(y)
\right]_x\,\,\delta_{m_y,0}\,\,\,
\frac{(-1)^{m_x}}{ m_x\Delta p_x}
\nonumber
\\[2mm]
&&\hskip-1cm
-e\left[\phantom{\frac11}
\Ev(\xv)+\frac{\Pv_\Mv}{m}\times\Bv(y)
\right]_y\,\,\delta_{m_x,0}\,\,\,
\frac{(-1)^{m_y}}{m_y\Delta p_y}
 -
(1-\delta_{\mv,0})
\frac {B_1 \hbar^2e}{12m}
\left(
\frac{(-1)^{m_x}}{ m_x\Delta p_x}
 \frac{(-1)^{m_y}}{ m_y\Delta p_y}
\frac{\partial}{\partial x}
+
\frac{2 (-1)^{m_y}}{ (m_y\Delta p_y)^2}
\frac{\partial}{\partial y}
\right)
\nonumber\\[2mm]
&&\hskip-1cm
-
(1-\delta_{m_y,0})
\frac {B_1 \hbar^2e}
{6m}
\frac{(-1)^{m_y}}{ (m_y\Delta p_y)^2}
\frac{\partial}{\partial y} -\delta_{\mv,0}
\frac {B_1}m
\frac e{12}
\frac{L_y^2}{12}\frac{\partial}{\partial y}
\left.
\phantom\int\right\}
f_w(\Pv_{\Mv-\mv},\xv)
\label{eqdiscrete}
\end{eqnarray}
\end{widetext}

Equation \eqn{IWE}  determined by the Fourier images of the EM fields reduces to  \eqn{eqdiscrete} with concrete kernel terms,  depending directly on the linear EM field components. The numerical challenges  are significantly reduced despite  the appearance of the terms of the alternating harmonic series.

In both the Boltzmann and the standard Wigner equation the integral kernel multiplies the  solution, which allows to present it  as a resolvent (Neumann) series~\cite{Dimov2007} determined by the consecutive operations of the kernel on the initial condition.  
The analysis of the corresponding resolvent series provides physical insights into the processes governing the evolution. 
In contrast, the right-hand side of \eqn{eqdiscrete} contains spatial derivatives of $f_w$, which complicate such analysis. Nevertheless, this can be done at the expense of further approximations of the equation. The idea for how to proceed comes from the long coherence limit of the equation, which will be discussed next. 

\subsection{\label{LCLL}Long Coherence Length Limit}

As discussed, the choice of a large, but finite $\Lv$ does not affect the behavior of the physical system inside $\Omega$.
 We will see that in this limit the
alternating harmonic series terms in \eqn{eqdiscrete} give rise to further derivatives on the
components of the momentum variable.
In a next step, we approximate the derivatives under the integral operator of the continuous equation by their finite difference representation.
This step allows to derive a Fredholm integral equation of the second kind which provides  a resolvent series expansion of the discrete momentum Wigner function.

The derivation of \eqn{eqdiscrete} involves mathematical operations which rely on the finite domain of integration. Thus, the derivation of the long coherence limit must begin from the
main equation ${\cal D}={\cal T}$, represented by expressions \eqn{Deafter}, \eqn{taucal}, and \eqn{taulin}, as discussed in the following. 

\subsubsection{First Magnetic Term}
We  begin with the analysis of the first magnetic term  in \eqn{Deafter}.
This term converges to the following expression in the limit $\Lv\to\infty$:
\begin{eqnarray}
&&\frac{B(y)e}{i\hbar m}
\int\limits_{-\Lv/2}^{\Lv/2}
\frac{d\sv}{\Lv}
\sum_{\mv=-\infty}^{\infty}
\left(
P_{M_x}s_y-P_{M_y}s_y
\right)
e^{-\frac i\hbar \mv \Delta\Pv \sv}
\label{firstmagnetic1_org}\\
&&
f_w\bigl(\Pv_{\Mv-\mv},\xv\bigr)
\longrightarrow
\frac{B(y)e}{m}\left(
P_x
\frac{\partial
}{\partial P_y}
-
P_y 
\frac{\partial
}{\partial P_x}\right)
f_w\bigl(\Pv,\xv\bigr)
\nonumber
\end{eqnarray}
Indeed, by recalling  that $\Delta \Pv=2\pi/\Lv$ and that
$\Pv_\mv=\mv\Delta\Pv$ the integrand function resembles the Riemann sum
of the "good" continuous and integrable function $f_w$  multiplied by a bounded function - the exponent.
The limit of the sum thus represents the corresponding regular integral, where
$P_{M_x}=M_x\Delta P_x\to P_x$, $P_{M_x}=M_x\Delta P_x\to P_x$.
It is straightforward to evaluate \eqn{firstmagnetic1_org} using integration by parts.
Details are given in Appendix~\ref{AAC}.

\subsubsection{Second Magnetic Term}
The second magnetic term involves more cumbersome calculations, because it contains both, higher order derivatives  and higher order products of the components of $\sv$. This requires consecutive steps of
integration by parts together with the assumption that $f_w$ vanishes at infinity along
with its derivatives. Nevertheless, the calculations are straightforward, giving rise to the
following limit:
\begin{eqnarray*}
&&-
\frac {B_1}m
\frac e{12}
\int\limits_{-L_y/2}^{L_y/2}\frac{ds_y}{L_y}
\\
&&
\sum_{m_y=-\infty}^{\infty}e^{-\frac i\hbar m_y \Delta P_y s_y}
s_y^2
\frac{\partial}{\partial x}f_w\bigl( M_x\Delta P_x, (M_y- m_y)\Delta P_y,\xv)
\\
&&
+\frac {B_1}m\frac e{12}
\int\limits_{-\Lv/2}^{\Lv/2}\frac{d\sv}{\Lv}
\sum_{\mv=-\infty}^{\infty}e^{-\frac i\hbar \mv \Delta \Pv\sv}
s_ys_x\frac{\partial}{\partial y}f_w\bigl(\Pv_{\Mv-\mv},\xv\bigr)
\\
&&
\longrightarrow
\frac {B_1\hbar^2}m\frac e{12}
\left(
\frac{\partial^2}{\partial P_y^2}\frac{\partial}{\partial x}
 -
\frac{\partial}{\partial P_x}\frac{\partial}{\partial P_y}
\frac{\partial}{\partial y}
\right)
f_w\bigl(P_x,P_y,\xv\bigr)
\end{eqnarray*}

\subsubsection{Electric Field Term}
The electric term in ${\cal T}$ is evaluated in the same way:
\begin{eqnarray*}
-
&&\left.
\frac {eE_xx}{i\hbar}
\int\limits_{-L_x/2}^{L_x/2}
\frac{ds_x}{L_x}
\sum\limits_{m_x=-\infty}^{\infty}
e^{-\frac i\hbar m_x\Delta P_x s_x}
s_x\right.f_w(\Pv_{\Mv-\mv}, \xv)
\\
&&-\left. \frac {eE_yy}{i\hbar}
\int\limits_{-L_y/2}^{L_y/2}
\frac{ds_y}{L_y}
\sum\limits_{m_y=-\infty}^{\infty}
e^{-\frac i\hbar m_y\Delta P_y s_y}
s_y
\right.f_w(\Pv_{\Mv-\mv}, \xv)\\
\longrightarrow
&&-
{eE_xx}
\frac{\partial f_w(P_x, P_y,\xv)}{\partial P_x}
-
{eE_yy}
\frac{\partial f_w(P_x, P_y,\xv)}{\partial P_y}
\end{eqnarray*}
By combining these results we obtain the continuous formulation of the
evolution equation for the Wigner function, see Appendix~\ref{AAE}.
The equation can be reformulated in a physically very informative way
\begin{eqnarray}
&&\left(\frac{\partial}{\partial t}
+
\frac{\Pv}{m}\cdot
\frac{\partial}{\partial \xv}
+\Fv(\Pv,\xv)
\cdot\frac{\partial}{\partial \Pv}\right)f_w\bigl(\Pv,\xv\bigr)
=\label{lform}
\\[2mm]
&&\frac {B_1\hbar^2}m\frac e{12}
\left(
\frac{\partial^2}{\partial P_y^2}\frac{\partial}{\partial x}
-
\frac{\partial}{\partial P_x}\frac{\partial}{\partial P_y}
\frac{\partial}{\partial y}\right)f_w\bigl(\Pv,\xv\bigr)
\nonumber
\end{eqnarray}

with the help of the Lorentz force $\Fv$: 

\begin{eqnarray*}
\Fv(\Pv,\xv)=
e\left(\Ev(\xv)+\frac{\Pv\times\Bv(y)}{m}\right) 
\end{eqnarray*}

The left-hand side of \eqn{lform} is the  Liouville operator of the  Boltzmann
equation, which determines the classical electron evolution. 
However, on the right-hand side the collision operator acting on $f_w$ is now
replaced by an operator depending on the magnetic field gradient. If the latter becomes zero
the equation consistently recovers the collisionless Boltzmann equation.
The right-hand side of the equation is thus responsible for all quantum effects in the
evolution. This analogy allows to interpret the quantum effects in the evolution
as a kind of scattering process. In contrast to the decoherence-causing stochastic scattering processes
unified in the Boltzmann  collision operator, the 
effect of the quantum-magnetic operator is not yet explored.
Indeed, an alternative analogy holds and is related to the Wigner potential operator, which, however, preserves the coherence. Furthermore, the quadratic magnetic-field-term has been neglected in \eqn{lform}.
The above considerations underline the importance of developing a numerical approach to solve the equation, which is discussed in the next section.

\subsection{\label{DIR} Discrete Integral Representation}
\subsubsection{Evolution Models}
The 
quantum-magnetic operator in \eqn{lform} involves third-order  
mixed position-momentum derivatives. In contrast, \eqn{eqdiscrete} contains only position
derivatives. Nevertheless, both model equations originate from the main equation ${\cal D}={\cal T}$:
The link becomes clear if one applies a finite difference scheme to the momentum derivatives
of the continuous equation $\partial /\partial \Pv\to \Delta/\Delta\Pv$.
With this step we return to  the finite coherence length description.
The link with  \eqn{eqdiscrete} is established by four straightforward
steps: 
(i) $\Pv$ is replaced back with the discrete momentum $\Pv_\Mv=\Mv\Delta \Pv$. 
(ii) The force term in \eqn{lform} is transferred to the right. 
 (iii) A finite difference scheme is adopted to represent 
the derivatives of the momentum components as linear combinations of terms of type
$f_w(\Pv_\Mv-\Qv,\xv)$, where $ \Qv$ is a vector with components $(Q_x\Delta P_x, Q_y\Delta P_y)$, $Q_{x,y}=\pm 0,1,2$. 
(iv) $f_w$ is represented as a sum $f_w(\Pv_{\Mv+\Qv},\xv)=\sum_{\mv}\delta_{\mv,\Qv}f(\Pv_{\Mv-\mv},\xv)$.
The particular expression depends on the used difference
scheme, however, in general, in the long coherence length limit the terms of the
harmonic series are replaced by Kronecker delta functions. 
This  simplifies the equation governing  the Wigner function and ensures an intuitive understanding  of the evolution, see Section~\ref{IR}. 
The numerical properties of the two models, \eqn{eqdiscrete} and \eqn{lform}, have yet to be investigated. Equation \eqn{eqdiscrete} is valid for any (finite) coherence length, while the  counterpart obtained from \eqn{lform} needs sufficiently long $\Lv$ to
ensure a "good" approximation of the derivatives $\partial /\partial \Pv\to \Delta/\Delta\Pv$.  

\subsubsection{\label{IR} Integral Transform}
Here, we further apply a finite difference scheme also to the spatial derivatives.
We use the characteristics of the zero force Liouville operator
\eq
\xv(t')=\xv - \int_{t'}^t\frac{\pv(\tau)}{m}d\tau\quad
\pv(t')=\Pv_\Mv \, , 
\label{Newton}
\eqend
which represents a free-streaming Newtonian trajectory.
The trajectory is initialized by the point $\Pv_\Mv, \xv, t$, while $t'<t$ is the 
running time. We consider the family of equations obtained
from \eqn{lform} (parametrized by $t'$) by replacing $\Pv, \xv$ by $\pv(t'),\xv(t')$ and write explicitly the
time dependence of $f_w$. 
With the help of \eqn{Newton}, the left-hand side 
can be written as a full time derivative:
\begin{eqnarray*}
&&\frac{d}{d t'}
\left( e^{-\int\limits^t_{t'}\gamma(\tau)d\tau}f_w\bigl(\pv(t'),\xv(t'),t'\bigr)\right)
= \frac {B_1\hbar^2e}{12m}e^{-\int\limits^t_{t'}\gamma(\tau)d\tau}
\\
&&
\left(
\frac{\Delta^3 f_w}{\Delta P_y^2\Delta x}
-\frac{\Delta^3f_w}{\Delta P_x\Delta P_y\Delta y}\right)\bigl(\pv(t'),\xv(t'),t'\bigr)
\\
&&\left(
-
\Fv
\cdot\frac{\Delta f_w}{\Delta \Pv}
+\gamma
f_w\right)\bigl(\pv(t'),\xv(t'),t'\bigr)
e^{-\int\limits^t_{t'}\gamma(\tau)d\tau}
\end{eqnarray*}
Here, we included the exponent of an auxiliary function $\gamma$~\footnote{The quantity $\gamma$ has to be determined based on numerical considerations. It is introduced in analogy with the treatment of the  Boltzmann and standard Wigner equations. In the former, $\gamma$ refers to the total out-scattering rate and with respect to the latter, it refers to the interaction rate with the Wigner potential. These choices for $\gamma$ provide a feasible and physically transparent numerical model.}.
Next, we consider the evolution of an initial condition $f_0$ specified at  time $t'=0$, and
integrate on $t'$ in the interval $(0,t)$:
\begin{eqnarray}
&&f_w\bigl(\Pv_\Mv,\xv,t\bigr)
= 
e^{-\int\limits^t_{0}\gamma(\tau)d\tau}f_0\bigl(\pv(0),\xv(0)\bigr)
\nonumber \\
&&
+
\int\limits_0^{t} dt'
e^{-\int\limits^t_{t'}\gamma(\tau)d\tau}
\left[
\frac {B_1\hbar^2e}{12m}
\left(
\frac{\Delta^3 f_w}{\Delta P_y^2\Delta x}
\bigl(\pv(t'),\xv(t'),t'\bigr)\right.\right.
\nonumber \\
&&
 \left.
-\frac{\Delta^3f_w}{\Delta P_x\Delta P_y\Delta y}\bigl(\pv(t'),\xv(t'),t'\bigr)
\right)
\label{final}
\\
&&
\left.
-\Fv
\cdot\frac{\Delta f_w}{\Delta \Pv}\bigl(\pv(t'),\xv(t'),t'\bigr)
+\gamma(t')
f_w\bigl(\pv(t'),\xv(t'),t'\bigr)\phantom\int\right]
\nonumber
\end{eqnarray}
The right-hand side contains linear combinations of the solution $f_w$ 
so that we obtain a Fredholm integral equation 
of a second kind. The phase space point $\Pv_\Mv,\xv$
and the time  $t$ initialize the trajectory on the right. Furthermore, the 
variable $t$ gives the upper limit of the $t'$ integral on the right, so that \eqn{final}
can be further specified as a  Volterra type  equation. It follows that the solution exist for
any evolution time $t$. It is represented by the resolvent series having terms
determined by the consecutive applications of the kernel on the  free term 
$e^{-\int\gamma d\tau}f_0$.                     
The initial condition  $f_0$
must thus be an admissible quantum state which contains the whole information about the physical system
and the involved spatial and momentum/energy characteristics.  
The terms in \eqn{final} are real quantities and so is  $f_w$. 
The function $\gamma$ provides a convenient way to analyze the role of the terms in \eqn{final}.
Indeed, if $\gamma$ is taken out of the square brackets, the product $\gamma e^{-\int\gamma d\tau}$
has the meaning of a probability distribution for any positive function $\gamma$.  Then the $t'$ integral in \eqn{final} becomes the expectation value of the random variable comprised by the terms enclosed in the square brackets.
Equation \eqn{final} has the same formal structure as the integral form of the classical Boltzmann equation~\cite{MixiBook2021}. The analysis of the resolvent expansion of the latter 
associates to the evolution consecutive processes of free flight and scattering events. The former proceeds over Newtonian trajectories and is interrupted by  scattering events with a probability given by $\gamma e^{-\int\gamma d\tau}$ where the classical $\gamma$, being the sum of all scattering rates, is called the out-scattering rate.
The latter are obtained with the help of the Fermi Golden Rule (FGR), giving rise to scattering events which are local in space, instantaneous, and only change the electron momentum.
The after-scattering state, which in classical mechanics is a phase-space and time point,  initializes the  trajectory for the next free flight and so on. This picture entirely emulates the physical model of evolution of the classical electron.  These considerations   offer a heuristic understanding of \eqn{final}.
The trajectories are determined by \eqn{Newton}, $\gamma$ plays the role of the  out-scattering rate, while
the terms in the square brackets can be interpreted as after-scattering states.  For example, $f_w(\Pv_\Mv- \Qv,\xv)$ can be considered as being obtained by the conservation relations inherent for the FGR, giving rise to a change in the momentum by $\Qv$ due to a scattering event with another particle, e.g., a phonon. 
Without the spatially non-local terms (accounting for the position derivatives) and with all signs switched to positive, \eqn{final} becomes an ordinary Boltzmann equation with fictitious, but physically admissible scattering mechanisms. 
The negative signs and the spatial non-locality is a manifest of the quantum character of \eqn{final}. Furthermore, the quantum state is a function in phase space and not a point as in the classical case.
However, the function can be represented by an ensemble of points, where the evolution is dictated by \eqn{Newton} and the resolvent series of \eqn{final}. This  heuristic picture of  the quantum transport process can be used as a base for developing stochastic particle methods for finding the solution of the quantum equation \eqn{final}.

\section{Discussion and Conclusions}
We use the WST of the von Neumann equation to develop a gauge-invariant
quantum mechanics theory in phase space.  The approach  relies on the FT 
of the EM field components. This imposes conditions about their behavior at infinity  or alternatively invokes the theory of generalized functions. This elegant division of the  mathematical analysis is based  on concepts and limits, which leave little space for 
a standard numerical treatment. 
This can be avoided on the expense of introducing a discrete momentum phase space. 
The approach is  motivated by the fact that quantum systems are bounded and often very small, nanoscale objects. This allows to apply a  FT based on  discrete momentum coordinates with spacing
determined by the coherence length. 
The derived equation \eqn{IWE} describes the evolution in terms of a discrete momentum
and, in principle, can be considered from a numerical point of view. 
However, for general EM fields, it contains multi-dimensional mathematical operations of summation and integration, being detrimental for practical application in numerical solution methods. To continue the  analysis and to gain  insights about this  gauge-invariant equation we need to assume a concrete shape of the  EM fields. It is suggested by the fact that for homogeneous (constant) EM conditions the equation reduces to its classical counterpart. It is thus relevant to consider the next
term in their Taylor expansion, namely  to consider a linear spatial dependence. 
This allows to reduce the numerical complexity by analytically performing  the $\tau$ integration.
The derived equation \eqn{eqdiscrete} can be further simplified to provide a heuristic information about
the peculiarities of the gauge-invariant model. We restrict to physical conditions which ignore the
nonlinear magnetic field dependence term in the kernel. 
A further approximation is suggested by the long coherence length limit of the equation. In this limit the summation turns into integration, however, the obtained expressions loose their meaning without a regularization
procedure, based on an exponential damping function. This procedure, formally used in the theory of distributions, is  physically justified by the fact that $\Omega$ is bounded. 
The result is the appearance of momentum  derivatives of the Wigner function, \eqn{lform}.
It can be reformulated as a Fredholm integral equation of a second kind if the derivatives are  approximated by using a  finite difference scheme. Actually there are two ways for doing this. The one, presented here, first approximates the derivatives and then uses the forceless trajectories \eqn{Newton}.
In this way, all operators appear on the right-hand side in \eqn{final} and their power can be compared as has been done in Appendix~\ref{AAA} for the particular physical setup.
Alternatively, the last line terms with the classical force in the continuous equation \eqn{CLEE} 
can be transferred to the left and one can use accelerated Newtonian trajectories to obtain the
integral form. The approximation of the derivatives results in a
Fredholm integral equation, where the kernel contains only quantum-related operators. This equation offers a numerical convenience because the  action of the  classical force operator is accounted for by the trajectory, being well-suited for numerical solution methods.

The assumption of a bounded domain  $\Omega$ for the evolution of a Schr{\"o}dinger state $\psi$ is a sufficient condition for the limiting procedure shown in Appendix~\ref{AAC}. It can be weakened by considering infinite domains
where the state tends to zero at infinity. However, this imposes an infinite coherence length which turns integrals into differential operators. Our experience with the linear case already shows that, the order of the derivatives rises with the power of $\sv$. Consequently, if further terms of the Taylor expansion of the EM fields are considered higher order derivatives are introduced in the equation: The order of the differential part of the so derived continuous evolution equation depends on the shape of the EM fields, which precludes a development of a general approach for finding the solution. The counterpart equation \eqn{IWE} has a well-defined differential part, but now the involved integrals depend on the EM fields. However, their replacement with  the corresponding Taylor expansion gives rise to well-defined integrals of polynomials of $\sv$ and $\tau$. Alternatively, one can use the corresponding Fourier representation of the EM fields and apply conventional analytical methods. 

We suggested a discrete momentum space formulation of the gauge-invariant Wigner theory and derived the general form of the evolution equation.  The established link between the discrete and the   continuous momentum equations for the important case of  linear EM fields reveals different aspects of the gauge-invariant theory. The numerical properties of the two formulations have yet to be investigated.

\section*{Acknowledgments}
The financial support by the Austrian Science Fund (FWF): P33609, P33151,
the Austrian Federal Ministry for Digital and Economic Affairs, 
the National Foundation for Research, Technology and Development, and 
the Christian Doppler Research Association
is gratefully acknowledged.

\appendix

\section{\label{AA0}Estimation of the Terms in ${\cal T}$}

The $\sv$ integration leads to the appearance of
the
alternating harmonic series terms in the expression for ${\cal T}$.
By using the components ${(M_x\Delta p_x, M_y\Delta p_y) }$ of $\Pv_\Mv$, we can rewrite ${\cal T}$
as follows:

\begin{eqnarray}
&&
{\cal T}=\frac{\partial}{\partial t}f_w(M_x\Delta p_x, M_y\Delta p_y,\xv) 
\label{tauend}\\&&
- {eE_xx}
\underset{m_x\ne 0}{\sum\limits_{m_x=-\infty}^{\infty}}
\frac{(-1)^{m_x}}{ m_x\Delta p_x}
f_w((M_x-m_x)\Delta p_x,M_y\Delta p_y,\xv)
\nonumber
\\
&&
- {eE_yy}
\underset{m_y\ne 0}{\sum\limits_{m_y=-\infty}^{\infty}}
\frac{(-1)^{m_y}}{m_y\Delta p_y}
f_w(M_x\Delta p_x,(M_y-m_y)\Delta p_y,\xv),
\nonumber
\end{eqnarray}

\section{\label{AAA}Estimation of the Terms in ${\cal D}$}
We consider the physical conditions in systems such as found in modern nanolelectronic structures which represent a broad class of evolution problems of practical relevance. They are featured by a direction of electron flow which
crosses  region(s) where the transport is dominated by quantum phenomena.
For example, let's consider a typical length of $20\, \rm nm$ and 
spatial variations in the order of $\Delta x \approx 1\, \rm nm$, which imposes the spacing between the points of the mesh discretization.
The actual structure dimensions motivate the choice of
a coherence length $L_c \approx 100\, \rm nm$, which corresponds to $100$ points sufficient to describe the spatial characteristics of the transport process.
This automatically determines the number of linked FT momentum points, as, in this case, there are 50 momentum values per positive/negative direction. 
The typical momentum value  $P_M = M \Delta P$ is then represented
by the middle point $M=25$. 
The coherence length then dictates
$\Delta P = \hbar \cdot 2\pi /L_c \simeq  7\cdot 10^{-27} \,$ \rm kg.m/s. With these values  the kinetic term
  $\frac{\Pv}{m}\cdot \frac{\partial}{\partial \xv}$ in equation (\ref{Deafter}), becomes   $10^{14} \div 10^{15}\, \rm  s^{-1}$
with $m$ chosen to be the mass of the free electron.
Actually $m^{-1}$ factors all of the evaluated terms and thus does not affect their relative magnitude, however, 
the units of $s^{-1}$ provide a heuristic measure about the time scale of action and the rate with which the corresponding operator modifies  the solution of the evolution equation.

The next three terms in (\ref{Deafter}) depend on the magnetic field.
We assume that $B(y)$ is defined by $B_0\simeq B_1L_c\simeq 1T$. The value of $s$ can be considered  comparable to the size of the here considered typical length of $ 20\, \rm nm$.

The first magnetic term of equation (\ref{Deafter}) differs from the
kinetic term by the factor $I=\frac{eB(y)}{\hbar} \cdot s \cdot \Delta x$
and is thus about two orders of magnitude smaller, having a magnitude of $10^{11} \, \rm s^{-1}$.
The third magnetic term is related to the second term by the same factor $I$:
It has a magnitude  of $10^{9} \, \rm s^{-1}$ and is, therefore, neglected:
In this way, only the linear terms with respect to $B_0$ and $B_1$ remain.

\section{\label{AAB} Linear Electromagnetic Fields Equation }
We present the  linear field variant of the finite coherence length equation \eqn{main};
The components of the vectors are given explicitly:
\begin{widetext}
\begin{eqnarray*} 
&&\left(\frac{\partial}{\partial t}
+
\frac{\Mv\Delta\Pv}{m}\cdot
\frac{\partial}{\partial \xv}\right)f_w(\Pv_\Mv,\mv)
=
-e\left[ {E_xx} + \frac{B(y) M_y\Delta P_y }{ m} \right]
\underset{m_x\ne 0}{\sum\limits_{m_x=-\infty}^{\infty}}
\frac{(-1)^{m_x}}{ m_x\Delta P_x}
f_w((M_x-m_x)\Delta P_x,M_y\Delta P_y,\xv)
\\
&&
\phantom{
\left(\frac{\partial}{\partial t}
+
\frac{\Mv\Delta\Pv}{m}\cdot
\frac{\partial}{\partial \xv}\right)f_w(\Mv\Delta\Pv,\mv)
=
}
-e\left[{E_yy}
-
\frac{B(y) M_x\Delta P_x }{ m}
\right]
\underset{m_y\ne 0}{\sum\limits_{m_y=-\infty}^{\infty}}
\frac{(-1)^{m_y}}{m_y\Delta P_y}
f_w(M_x\Delta P_x,(M_y-m_y)\Delta P_x,\xv)
\\
&&
\phantom{
\left(\frac{\partial}{\partial t}
+
\frac{\Mv\Delta\Pv}{m}\cdot
\frac{\partial}{\partial \xv}\right)f_w(\Mv\Delta\Pv,\mv)
=
}
 -
 \sum\limits_{\underset{m_x,m_y\ne 0}{ m_x,m_y=-\infty}}^\infty
\frac {B_1 \hbar^2e}{12m}
\left(
\frac{(-1)^{m_x}}{ m_x\Delta P_x}
 \frac{(-1)^{m_y}}{ m_y\Delta P_y}
\frac{\partial}{\partial x}
+
\frac{2 (-1)^{m_y}}{ (m_y\Delta P_y)^2}
\frac{\partial}{\partial y}
\right)
f_w(\Pv_{\Mv-\mv},\xv)
\nonumber\\
&&\hskip3.5cm
-\sum\limits_{\underset{m_y\ne 0}{m_y=-\infty}}^\infty
\frac {B_1 \hbar^2e}
{6m}
\frac{(-1)^{m_y}}{ (m_y\Delta P_y)^2}
\frac{\partial}{\partial y}
f_w(M_x\Delta P_x, (M_y-m_y)\Delta P_y,\xv)
-\frac {B_1}m
\frac e{12}
\frac{L_y^2}{12}
\frac{\partial}{\partial y}
f_w(\Pv_\Mv,\xv)
\end{eqnarray*}
\end{widetext}
We can reformulate the equation in a vector form by
observing that the vector product of
the terms in the square brackets are the components of the vector
$
e\left(\Ev(\xv)+\frac{\Pv_\Mv}{m}\times\Bv(y)\right)
$, giving rise to \eqn{eqdiscrete}.

\section{\label{AAC} Long Coherence Length Limit }
We consider the first term in the bracket of \eqn{firstmagnetic1_org}, which can be reformulated with the help of \eqn{complete}. After performing the $s_x$ integration  and introducing  the constant $C=\frac{B(y)e}{i\hbar m}$, we obtain
\begin{eqnarray*}
&&C
\int\limits_{-L_y/2}^{L_y/2}\frac{ds_y}{2\pi\hbar}
\\
&&\sum_{m_y=-\infty}^{\infty}
P_x 
s_y
e^{-\frac i\hbar m_y \Delta P_y s_y}f_w\bigl( P_y - m_y\Delta P_y,\cdot\bigr)
\frac{2\pi\hbar}{L_y}  \, ,
\end{eqnarray*}
where $P_x=M_x\Delta P_x$, $P_y=M_y\Delta P_y$,
and the dot ($\cdot$) denotes variables which are irrelevant for the here considered derivations.
By denoting $\xi_y=m_y\Delta P_y$, we observe
that the last factor in the above expression
is $\Delta P_y$ so that in the limit $L_y\to\infty$ 
the sum tends to the integral:
\eq
I=
CP_x 
\int\limits_{-\infty}^{\infty}\frac{ds_y}{2\pi\hbar}
\int_{-\infty}^{\infty}d\xi_y
s_y
e^{-\frac i\hbar \xi_y s_y}f_w\bigl( P_y - \xi_y,\cdot\bigr)
\eqend
This is devoid of meaning without a  proof of the convergence of the $s_y$ integral.
However, there is a mighty approach for regularization of such expressions. 
It redefines the integral by assigning  the function $e^{-\alpha |s_y|} $ in the limit  $I=\lim_{\alpha\to 0}I_\alpha$ to the integrand~\footnote{In our case this procedure inserts  the information that the spatial support of the physical system is finite.}:
\begin{eqnarray}
&&
I_{\alpha}=
CP_x 
\int\limits_{-\infty}^{\infty}\frac{ds_y}{2\pi\hbar}e^{-\alpha|s_y|}
\nonumber\\
&&
\int_{-\infty}^{\infty}d\xi_y
\frac{\hbar}{-i}
\left(
\frac{\partial}{\partial \xi_y}
e^{-\frac i\hbar \xi_y s_y}\right)f_w\bigl( P_y - \xi_y,\cdot\bigr)
\nonumber \\
&&
=
\frac{\hbar CP_x}{-i}
\int\limits_{-\infty}^{\infty}\frac{ds_y}{2\pi\hbar}e^{-\alpha|s_y|}
e^{-\frac i\hbar \xi_y s_y}f_w\bigl( P_y - \xi_y,\cdot\bigr)\bigg|_{-\infty}^{\infty}-
\nonumber \\
&&
\frac{\hbar CP_x}{-i2\pi\hbar}
\int\limits_{-\infty}^{\infty}d\xi_y
 \int\limits_{0}^{\infty}ds_y
\left(e^{-\frac i\hbar (\xi_y-i\hbar\alpha )s_y}+
e^{\frac i\hbar (\xi_y+i\hbar\alpha )s_y}
\right)
\nonumber \\&&
\frac{\partial f_w\bigl( P_y - \xi_y,\cdot\bigr)}{\partial \xi_y}
=
\label{IbyParts}\\
&&
-\frac{\hbar CP_x}{2\pi}
\int\limits_{-\infty}^{\infty}d\xi_y
\left(
\frac{1}{ \xi_y+i\hbar\alpha }
-\frac{1}{\xi_y-i\hbar\alpha }
\right)
\frac{\partial f_w\bigl( P_y - \xi_y,\cdot\bigr)}{\partial \xi_y}
\nonumber
\end{eqnarray}
Here, we can apply  the Sokhotski formula
\eq
\lim_{\alpha\to 0}\int\limits_{-\infty}^{\infty}\frac{\phi(x)}{x\pm i\epsilon}=\mp i\pi\phi(0)+
\lim_{\alpha\to 0}\left(\int\limits_{-\infty}^{-\alpha}+
\int\limits^{\infty}_{\alpha}\right)\frac{\phi(x)}{x} 
\eqend
 which can be symbolically written in terms of a delta function $\delta$ and a Cauchy principal value
$VP$ as:
\eq
\lim_{\alpha\to 0}\frac{1}{x\pm i\epsilon}=\mp i\pi\delta(x)+VP(\frac1x)
\eqend
 In the formal theory of distributions $\phi$ is  
assumed to belong to the class $D$ of infinitely differentiable
functions with a compact support. For our case it is sufficient that the first momentum derivative of $f_w$
is continuous with a compact support function.
Therefore, the limit of the bracket in \eqn{IbyParts} gives $-i2\pi\delta(\xi_y)$ and thus
$$
I=\frac{B(y)e}{m}
P_x 
\frac{\partial
f_w\bigl( P_y,\cdot\bigr)}{\partial P_y}
$$
and finally 
\begin{widetext}
\eq
\frac{B(y)e}{i\hbar m}
\int\limits_{-L_y/2}^{L_y/2}\frac{ds_y}{2\pi\hbar}
\sum_{m_y=-\infty}^{\infty}
P_x 
s_y
e^{-\frac i\hbar m_y \Delta P_y s_y}f_w\bigl( P_y - m_y\Delta P_y,\cdot\bigr)
\frac{2\pi\hbar}{L_y}
\longrightarrow
\frac{B(y)e}{m}
P_x
\frac{\partial
f_w\bigl( P_y,\cdot\bigr)}{\partial P_y} \, .
\label{firstmagneticI}
\eqend
In the same way we evaluate the second term:
\eq
-\frac{B(y)e}{i\hbar m}
\int\limits_{-L_x/2}^{L_x/2}\frac{ds_x}{2\pi\hbar}
\sum_{m_x=-\infty}^{\infty}
P_y 
s_x
e^{-\frac i\hbar m_x \Delta P_x s_x}f_w\bigl( P_x - m_x\Delta P_x,\cdot\bigr)
\frac{2\pi\hbar}{L_x}
\longrightarrow
-\frac{B(y)e}{m}
P_y 
\frac{\partial
f_w\bigl( P_x,\cdot\bigr)}{\partial P_x}
\label{firstmagneticII}
\eqend
\end{widetext}

\section{\label{AAE} Continuous Limit of the Evolution Equation}
By combining the evaluated terms we conclude that the long coherence length limit 
of the approximated  equation ${\cal T}={\cal D}$
gives rise to the following continuous formulation of the evolution
equation for the linear EM Wigner function:
\begin{eqnarray}
&&\left(\frac{\partial}{\partial t}
+
\frac{\Pv}{m}\cdot
\frac{\partial}{\partial \xv}\right)f_w\bigl(\Pv,\xv\bigr)
=
\nonumber\\[2mm]
&&
\left[
\frac{B(y)e}{m}\left(
P_x
\frac{\partial
}{\partial P_y}
-
P_y 
\frac{\partial
}{\partial P_x}\right)\right.
\nonumber\\[2mm]
&&+\frac {B_1\hbar^2}m\frac e{12}\left(
\frac{\partial^2}{\partial P_y^2}\frac{\partial}{\partial x}
-
\frac{\partial}{\partial P_x}\frac{\partial}{\partial P_y}
\frac{\partial}{\partial y}
\right)
\nonumber\\[2mm]
&&\left.-
{eE_xx}
\frac{\partial 
}{\partial P_x}
-
{eE_yy}
\frac{\partial 
}{\partial P_y}
\right]f_w\bigl(\Pv,\xv\bigr)
\label{CLEE}
\end{eqnarray}
This result can be further processed to obtain a form which enlightens the
involved physics: 
The terms in the second and forth row
can be unified to give the Lorentz force
$
\Fv(\Pv,\xv)=
e\left(\Ev(\xv)+\frac{\Pv\times\Bv(y)}{m}\right),
$
where we recall that the vectors in the product are defined as 
$\Pv=(P_x,P_y,0), \Bv=(0,0,B(y))$. Then the term with $\Fv$ can be transferred 
to the left to give \eqn{lform}.
\clearpage

\bibliographystyle{apsrev4-2}

\providecommand{\noopsort}[1]{}\providecommand{\singleletter}[1]{#1}%

\end{document}